\documentclass[aps,preprintnumbers,eqsecnum,amsmath,amssymb,showpacs,nofootinbib]{revtex4}
\usepackage{graphicx}% Include figure files
\usepackage{dcolumn}% Align table columns on decimal point
\usepackage{bm}% bold math
\usepackage{epsfig}

\begin{document}
%\preprint{HEPHY-PUB 889/10}
%\preprint{}
\title{\boldmath OPE, charm-quark mass, and decay constants \\ of $D$ and $D_s$ mesons from QCD sum rules}
\author{Wolfgang Lucha$^{a}$, Dmitri Melikhov$^{a,b,c}$, and Silvano Simula$^{d}$}
\affiliation{
$^a$HEPHY, Austrian Academy of Sciences, Nikolsdorfergasse 18, A-1050, Vienna, Austria\\
$^b$SINP, Moscow State University, 119991, Moscow, Russia\\
$^c$Faculty of Physics, University of Vienna, Boltzmanngasse 5, A-1090 Vienna, Austria\\
$^d$INFN, Sezione di Roma III, Via della Vasca Navale 84, I-00146, Roma, Italy}
\date{\today}
\begin{abstract}
We present a sum-rule extraction of the decay constants of the charmed mesons $D$ and $D_s$ from the two-point correlator of
pseudoscalar currents. First, we compare the perturbative
expansion for the correlator and the decay constant performed in
terms of the pole and the running $\overline{\rm MS}$ masses of
the charm quark. The perturbative expansion in terms of the pole
mass shows no signs of convergence whereas reorganizing this very
expansion in terms of the $\overline{\rm MS}$ mass leads to a
distinct hierarchy of the perturbative expansion. Furthermore, the
decay constants extracted from the pole-mass correlator turn out
to be considerably smaller than those obtained by means of the
$\overline{\rm MS}$-mass correlator.~Second, making use of the OPE
in terms of the $\overline{\rm MS}$ mass, we determine the decay
constants of both $D$ and $D_s$ mesons with an emphasis on the
uncertainties in these quantities related both to the input QCD
parameters and to the limited accuracy of the method of sum rules.
\end{abstract}
\pacs{11.55.Hx, 12.38.Lg, 03.65.Ge}
\maketitle

\section{Introduction}
The extraction of the decay constants of ground-state heavy
pseudoscalar mesons within the method of QCD sum rules
\cite{svz,aliev} poses a complicated problem. First, one derives
an operator product expansion (OPE) for the correlation function
\begin{eqnarray}
\label{1.1}
\label{Pi_QCD} \Pi(p^2)=i \int d^4x\, e^{ipx}\langle 0|T\left(j_5(x)j^\dagger_5(0)\right)| 0\rangle
\end{eqnarray}
of two pseudoscalar heavy-light currents
\begin{eqnarray}
j_5(x)=(m_Q+m)\bar q(x) i\gamma_5 Q(x).
\end{eqnarray}
Second, one considers the sum rule for this correlator. The sum
rule is nothing but the expression of the fact that the
representation of the Borelized correlator (\ref{1.1}),
$\Pi(p^2)\to \Pi(\tau)$, in the language of the intermediate
hadron states~is equal to the OPE for this correlator:
\begin{eqnarray}
\label{pitau} \Pi(\tau)=f_Q^2 M_Q^4 e^{-M_Q^2\tau}+
\int\limits_{s_{\rm phys}}^{\infty}ds\, e^{-s \tau} \rho_{\rm
hadr}(s)
=
\int\limits^\infty_{(m_Q+m)^2}ds\, e^{-s\tau}\rho_{\rm
pert}(s,\mu) + \Pi_{\rm power}(\tau,\mu).
\end{eqnarray}
Here, $M_Q$ denotes the mass of the pseudoscalar meson $P_Q$
containing the heavy quark $Q$ while $f_Q$ is its decay constant:
\begin{eqnarray}
\label{decay_constant} (m_Q+m) \langle 0 |\bar q i\gamma_5 Q| P_Q
\rangle = f_Q M_Q^2.
\end{eqnarray}
For the correlator (\ref{1.1}), $s_{\rm phys}=(M_{V_Q}+M_\pi)^2$
is the physical continuum threshold, $M_{V_Q}$ being the mass of
the vector meson containing $Q$. Obviously, for large values of
$\tau$ the contribution of the excited states decreases faster
than the ground-state contribution and therefore $\Pi(\tau)$ is
dominated by the ground state.

The perturbative spectral density is obtained in the form of an
expansion in terms of the strong coupling $\alpha_{\rm s}(\mu)$:
\begin{eqnarray}
\label{rhopert} \rho_{\rm
pert}(s,\mu)=\rho^{(0)}(s)+\frac{\alpha_{\rm
s}(\mu)}{\pi}\rho^{(1)}(s)+ \left(\frac{\alpha_{\rm
s}(\mu)}{\pi}\right)^2\rho^{(2)}(s)+\cdots.
\end{eqnarray}
Clearly, the correlator (\ref{1.1}) does not depend on the
renormalization scale $\mu$; however, both the perturbative
expansion truncated at fixed order in $\alpha_{\rm s}$ and the
truncated power corrections $\Pi_{\rm power}(\tau,\mu)$ given in
terms of the condensates and the radiative corrections to the
latter depend on $\mu$. Moreover, the relative magnitudes of the
lowest-order contributions strongly depend on the choice of the
renormalization scheme/scale.

Unfortunately, the truncated OPE allows one to calculate the
correlator only at not sufficiently large $\tau$, such that the
excited states give a sizable contribution to $\Pi(\tau)$ in the
corresponding $\tau$-range. In principle, the physical spectral
density above the threshold might be measured experimentally; in
practice, however, it is unknown. Therefore, one adopts the
concept of duality to relate the contribution of the excited
hadron states to the perturbative contribution: perturbative QCD
spectral density $\rho_{\rm pert}(s)$ and hadron spectral density
$\rho_{\rm hadr}(s)$ are close to each other at large values of
$s$; thus, for sufficiently large values of the parameter $\bar
s$, (far) above the resonance region, one has the duality relation
\begin{eqnarray}
\label{duality1} \int\limits_{\bar s}^{\infty} ds\, e^{-s \tau}
\rho_{\rm hadr}(s) = \int\limits_{\bar s}^{\infty}ds\, e^{-s
\tau}\rho_{\rm pert}(s).
\end{eqnarray}
In order to express the excited-state contribution by the
perturbative contribution, we need to extend this relationship
down to the value of the hadronic threshold $s_{\rm phys}$.
However, one has to be careful: the spectral densities $\rho_{\rm
pert}(s)$ and $\rho_{\rm hadr}(s)$ are obviously different in the
region near $s_{\rm phys}$. Therefore, one finds
\begin{eqnarray}
\label{duality} \int\limits_{s_{\rm phys}}^{\infty} ds\, e^{-s
\tau} \rho_{\rm hadr}(s) = \int\limits_{s_{\rm
eff}(\tau)}^{\infty} ds\, e^{-s \tau} \rho_{\rm pert}(s),
\end{eqnarray}
where $s_{\rm eff}(\tau)$ is different from the physical threshold
$s_{\rm phys}$. A crucial (albeit rather obvious) observation is
that,~for~the same reason which causes $s_{\rm eff}(\tau)\ne
s_{\rm phys}$, $s_{\rm eff}(\tau)$ {\em has to be\/} a function of
the parameter $\tau$ to render relation (\ref{duality}) exact.

By virtue of (\ref{duality}) we may rewrite the sum rule
(\ref{pitau}) as
\begin{eqnarray}
\label{sr} f_Q^2 M_Q^4 e^{-M_Q^2\tau}= \int\limits^{s_{\rm
eff}(\tau)}_{(m_Q+m)^2} ds\, e^{-s\tau}\rho_{\rm pert}(s,\mu) +
\Pi_{\rm power}(\tau,\mu) \equiv \Pi_{\rm dual}(\tau,s_{\rm
eff}(\tau)).
\end{eqnarray}
We refer to the right-hand side of this equation as the {\it dual 
correlator}. Evidently, even if the QCD inputs $\rho_{\rm
pert}(s,\mu)$ and $\Pi_{\rm power}(\tau,\mu)$ are well-known, the
extraction of the decay constant requires a further criterion for
fixing the effective continuum threshold $s_{\rm eff}(\tau)$.

Noteworthy, Eq.~(\ref{sr}) offers another way to convince oneself
that $s_{\rm eff}(\tau)$ must be a function of $\tau$. In fact,
the~log~slope on the left-hand side of (\ref{sr}) is independent
of $\tau$ and is equal to $M_Q^2$ (which may be exactly known from
experiment). Consequently, to guarantee the same $\tau$-behaviour
on the right-hand side of (\ref{sr}), the effective threshold must
be, in general, a function of $\tau$. In the literature the
approximation of the threshold by some constant $s_0$ independent of $\tau$ is widely used. 
The corresponding dual correlator, $\Pi_{\rm dual}(\tau, s_0)$,
should therefore lead to the presence of a contamination of excited states on the left-hand side of Eq.~(\ref{sr}). 
In principle, one may develop models for excited states in order~to estimate (and subsequently remove) 
such a contamination. It is, however, clear that ultimately the same effect can~be equivalently
reached by considering an explicit $\tau$-dependence of the effective continuum threshold.

The {\it exact effective continuum threshold\/}---corresponding to
exact values of the hadron mass and the decay constant on the
left-hand side---is, of course, not known. Therefore, the actual
extraction of hadron parameters from a sum~rule consists in
attempting (i) to find some reasonable approximation to the exact
threshold and (ii) to control the accuracy of such an
approximation. We stress again that the use of a $\tau$-dependent
threshold is expected to improve the reliability of the extraction
of the hadron parameter considered compared with the {\it standard} procedure 
of assuming a constant, $\tau$-independent threshold.

Let us now look in detail at each step of the sum-rule calculation
of the decay constant, starting with the OPE for the correlator.

%\vspace{-.5cm}
\section{OPE and heavy-quark mass}
We use the perturbative spectral density $\rho_{\rm pert}(s)$ calculated in \cite{chetyrkin} 
to three-loop accuracy in terms of the pole mass~of the heavy quark. 
The pole mass has been used in most of the sum-rule analyses since the pioneering work \cite{aliev}. 
An alternative option is to reorganize the perturbative expansion in terms of the running
$\overline{\rm MS}$ mass \cite{jamin}. Since the correlator is known to $\alpha_s^2$-accuracy, 
the relationship between pole and $\overline{\rm MS}$ mass to the same accuracy is used.~Explicit
expressions for the perturbative spectral densities and power corrections may be found in \cite{chetyrkin,jamin} 
and are not given~here.

\begin{figure}[!h]
\begin{tabular}{cc}
\includegraphics[width=6.5cm]{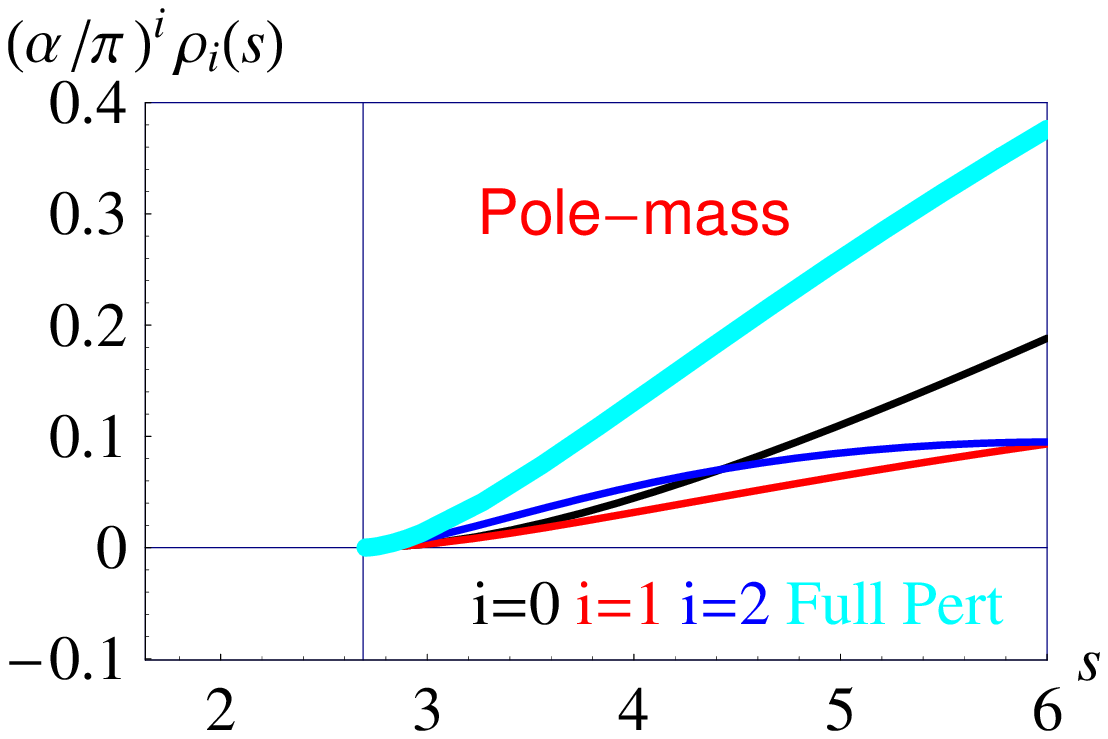}
&
\includegraphics[width=6.5cm]{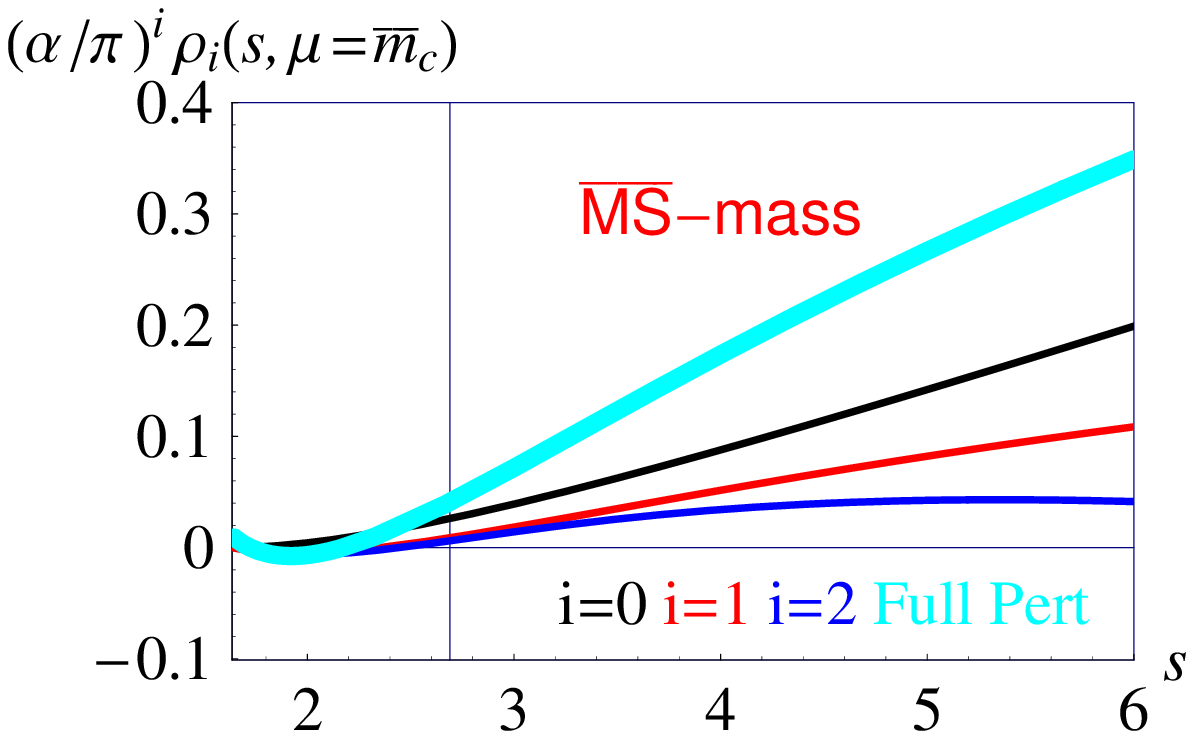}
\\
\includegraphics[width=6.5cm]{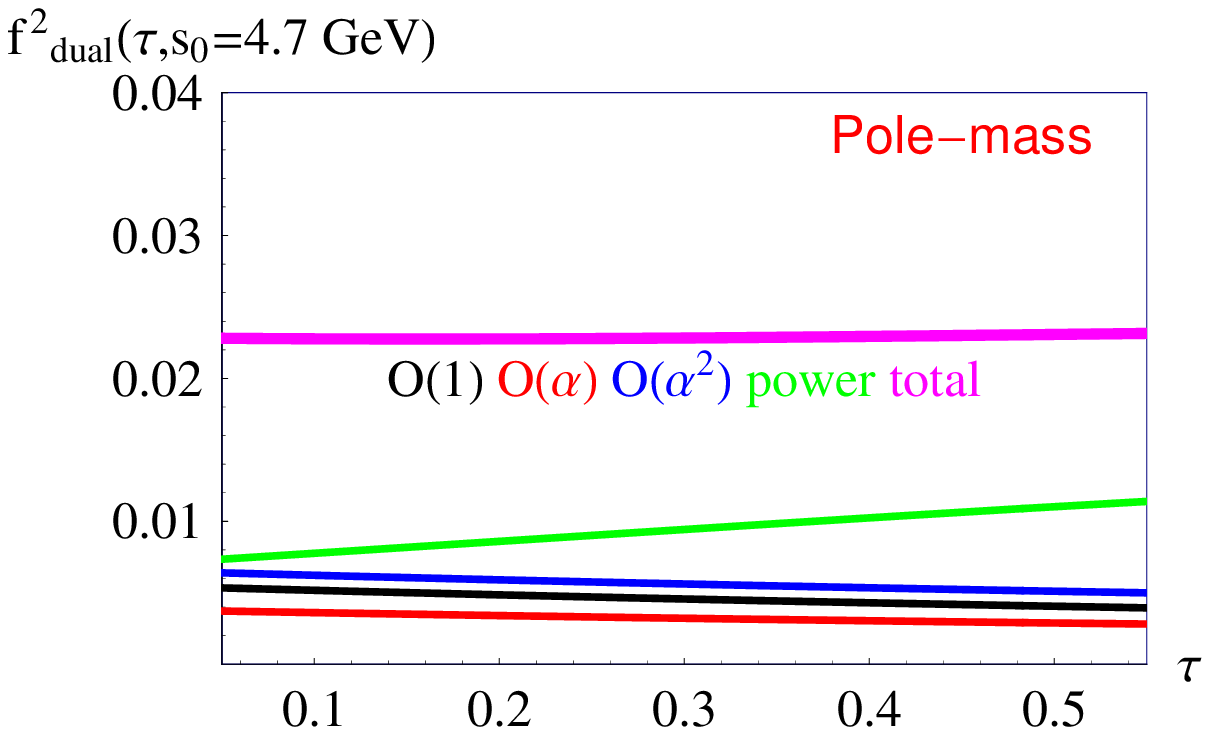}
&
\includegraphics[width=6.5cm]{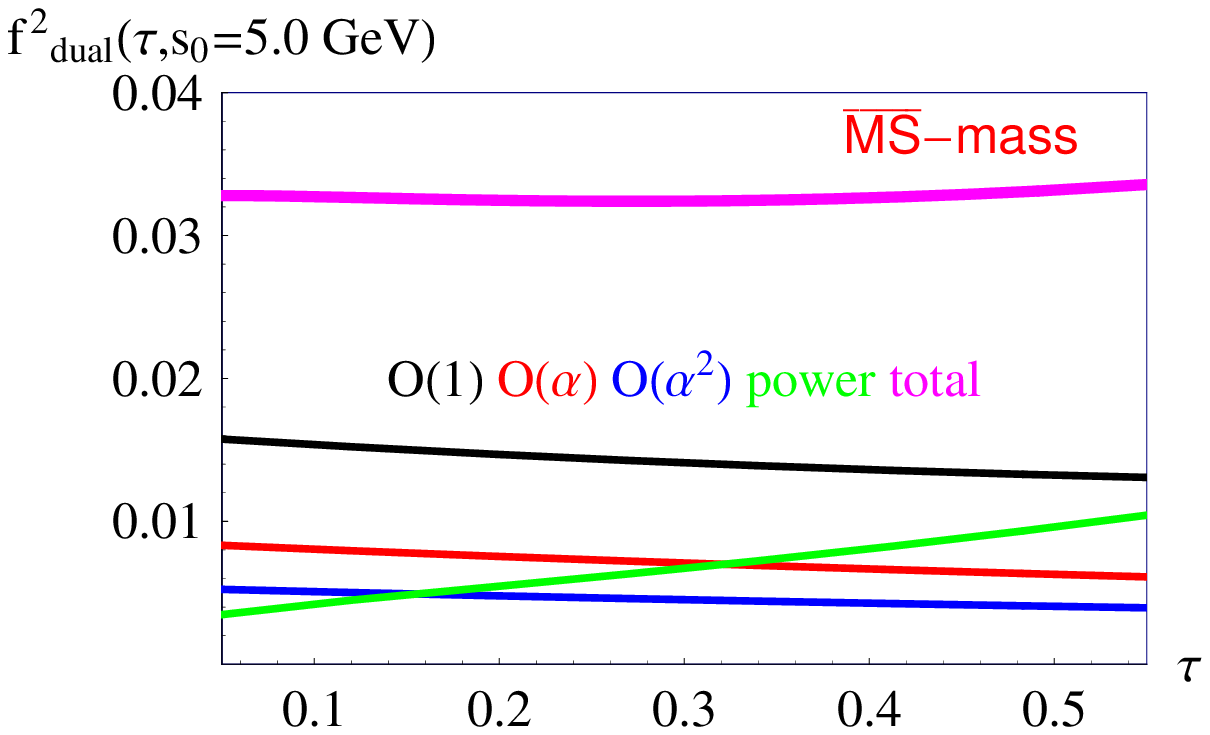}
\\
\end{tabular}
\caption{\label{Plot:1} OPE calculated in terms of the pole mass
(left) and the $\overline{\rm MS}$ mass (right) of the $c$ quark.
First line: spectral densities; second line: corresponding
sum-rule estimates for $f_D$. A constant effective continuum
threshold $s_0$ is fixed in each case~separately by requiring
``maximal stability'' of the extracted decay constant. As the
result, $s_0$ turns out to be different in the two schemes.}
\end{figure}

Figure~\ref{Plot:1} shows the perturbative spectral densities and
the sum-rule estimates for $f_D$ arising from (\ref{sr}) for our
two choices of $m_c$: the pole mass $m_{\rm c,pole}$ and the
running $\overline{\rm MS}$ mass $\overline{m}_c(\mu)$. The relevant OPE parameters are
\begin{align}
%\begin{eqnarray}
\label{Table:1} 
&\overline{m}_c(\overline{m}_c)=(1.279\pm 0.013)\;{\rm GeV},\quad 
m(2\;{\rm GeV})=(3.5\pm 0.5)\;{\rm MeV},\quad 
m_s(2\;{\rm GeV})=(100\pm 10)\;{\rm MeV},
\nonumber\\
&\alpha_S(M_Z)=0.1176\pm 0.0020,
\\ 
&\langle\bar qq\rangle(2\;{\rm GeV})=-((267\pm 17)\;{\rm MeV})^3,\quad 
\langle\bar ss\rangle(2\;{\rm GeV})/\langle\bar qq\rangle(2\;{\rm GeV})=0.8\pm 0.3,
\quad\left\langle\frac{\alpha_s}{\pi}GG\right\rangle=(0.024\pm 0.012)\;{\rm GeV}^4.
\nonumber
\end{align}
%\end{eqnarray}
We employ a recent determination \cite{mb} of $\overline{m}_c(\overline{m}_c)$. 
The corresponding pole mass recalculated from the $O(\alpha_s^2)$ relation between
$\overline{m}_c$ and $m_{\rm c,pole}$ is
\begin{eqnarray}
m_{\rm c,pole}=1.682\;{\rm GeV}.
\end{eqnarray}
The sum-rule estimates shown in Fig.~\ref{Plot:1} are obtained for
a $\tau$-independent effective threshold $s_0$. Its values,
which~prove to be different for the pole-mass OPE and the
$\overline{\rm MS}$-mass OPE, are found by requiring maximal
stability of the extracted decay constant. Obviously, for
heavy-light correlators and the resulting decay constants it makes
a very big difference which precise scheme for the heavy-quark
mass is employed.

Several lessons should be learnt from these plots:

\noindent (i) The perturbative expansion for the decay constant in
terms of the pole mass shows no signs of convergence; each~of the
terms---LO, NLO, NNLO---gives contributions of similar size.
Therefore, there is no reason to expect higher~orders to give
smaller contributions.

\noindent (ii) Reorganizing the perturbative series in terms of
the $\overline{\rm MS}$ mass of the heavy quark leads to a clear
hierarchy of the perturbative contributions.
%Notice however, that also in this case the situation is not perfect:
%the full spectral density -- which is a positive-definite function -- is negative at small values of $s$.
%This is an artifact of truncation and indicates that the contributions of higher-order terms are non-negligible.

\noindent (iii) The absolute value of the decay constant extracted
from the pole-mass OPE ($f_D=150$ MeV) proves to be considerably smaller
than that from the $\overline{\rm MS}$ scheme ($f_D=180$ MeV). Let us
emphasize that, nevertheless, in both cases the decay~constant
exhibits perfect stability in a wide range of the Borel parameter
$\tau$! Thus we emphasize again that mere Borel~stability~is by
far not sufficient to guarantee the reliability of the sum-rule
extraction of bound-state parameters. We have already observed
this feature in several examples in quantum mechanics
\cite{lms_1}.

Because of the obvious problems with the pole-mass OPE for the
correlator, we shall make use of the OPE in terms of the running
$\overline{\rm MS}$ mass for our extraction of the decay
constants. Hereafter, the quark masses $m_Q$ and $m$, and the
strong coupling $\alpha_s$ denote the $\overline{\rm MS}$ running
quantities at the scale $\mu$.

\section{Extraction of the decay constant}
In order to determine the heavy-meson decay constant $f_Q$ from
the OPE, we must execute the following two steps.

\subsubsection{The Borel window}
First, we must fix our working $\tau$-window where, on the one
hand, the OPE gives a sufficiently accurate description~of the
exact correlator (i.e., all higher-order radiative and power
corrections are small) and, on the other hand, the ground state
gives a ``sizable'' contribution to the correlator. Since the
radiative corrections to the condensates increase rather fast with
$\tau$, it is preferable to stay at the lowest possible values of
$\tau$. We shall therefore fix the window by the following
criteria \cite{lms_new,lms_qcdvsqm}: (a) In the window, power
corrections $\Pi_{\rm power}(\tau)$ should not exceed 30\% of the
dual correlator $\Pi_{\rm dual}(\tau,s_0)$. This restricts the
upper boundary of the $\tau$-window. The ground-state contribution
to the correlator at this value of~$\tau$ comprises about 50\% of
the correlator. (b) The lower boundary of the $\tau$-window is
fixed by the requirement that the ground-state contribution does
not fall below 10\%.

\subsubsection{The effective continuum threshold}
Second, we must define a criterion how to determine $s_{\rm
eff}(\tau)$. The corresponding algorithm has been formulated in
our recent works \cite{lms_new,lms_qcdvsqm} and was shown to
provide a good extraction of the ground-state parameters in
\mbox{quantum-mechanical} potential models.

Let us introduce the {\em dual invariant mass\/} $M_{\rm dual}$
and the {\em dual decay constant\/} $f_{\rm dual}$ by the
relations
\begin{eqnarray}
\label{mdual} M_{\rm dual}^2(\tau) \equiv -\frac{d}{d\tau}\log
\Pi_{\rm dual}(\tau, s_{\rm eff}(\tau)),\qquad
%\\
\label{fdual}
f_{\rm dual}^2(\tau)
\equiv
M_Q^{-4} e^{M_Q^2\tau}\Pi_{\rm dual}(\tau, s_{\rm eff}(\tau)).
\end{eqnarray}
For a properly constructed $\Pi_{\rm dual}(\tau, s_{\rm
eff}(\tau))$, this dual mass should coincide with the actual mass
of the ground state.~So, if the ground-state mass is known, any
deviation of the dual mass from the actual mass of the ground
state yields an indication of the contamination of the dual
correlator by excited states.

Assuming some particular functional form of the effective
threshold and requiring the least deviation of the dual~mass
(\ref{mdual}) from the actual mass in the $\tau$-window entails a
variational solution for the effective threshold; as soon as the
latter has been fixed, (\ref{fdual}) yields the decay constant.
The standard assumption for the effective threshold is a
$\tau$-independent constant. In addition to this approximation, we
also consider polynomials in $\tau$.

Our algorithm for the extraction of $f_Q$ makes use of the
knowledge of the true $P_Q$-meson mass $M_Q$. This algorithm,
developed in our previous works and proven to work well for
different correlators in the potential model, is very simple: we
consider the set of $\tau$-dependent Ans\"atze for the effective
continuum threshold
\begin{eqnarray}
\label{zeff} s^{(n)}_{\rm eff}(\tau)=
\sum\limits_{j=0}^{n}s_j^{(n)}\tau^{j}.
\end{eqnarray}
We fix the parameters on the right-hand side of (\ref{zeff}) as
follows: we compute the dual mass squared according to
(\ref{mdual}) for the $\tau$-dependent $s_{\rm eff}(\tau)$ in
(\ref{zeff}). We then evaluate $M^2_{\rm dual}(\tau)$ at several
values of $\tau = \tau_i$ ($i = 1,2,\dots,N$,~where~$N$~can be
taken arbitrary large) chosen uniformly in the window. Finally, we
minimize the squared difference between $M^2_{\rm dual}$ and the
known value $M^2_B$:
\begin{eqnarray}
\label{chisq}
\chi^2 \equiv \frac{1}{N} \sum_{i = 1}^{N} \left[ M^2_{\rm dual}(\tau_i) - M_Q^2 \right]^2.
\end{eqnarray}
This gives us the coefficients $s_j^{(n)}$ of the effective
continuum threshold. As soon as the latter is fixed, it is
straightforward to calculate the decay constant.

The results presented below indicate that accounting for the
$\tau$-dependence of the effective threshold yields a visible
improvement compared with the usual assumption of a
$\tau$-independent quantity in the following respect: it leads to
a much better stability of the dual mass calculated for a dual
correlator, which is tantamount to a better isolation~of~the
ground-state contribution.

Still, by trying different Ans\"atze for the effective continuum
threshold, one obtains different estimates for the decay constant.
We discuss the interpretation of these results in connection with
the systematic uncertainties of the method~of sum rules.

\subsubsection{Uncertainties in the extracted decay constant}
Clearly, the extracted value of the decay constant is sensitive to
the precise values of the OPE parameters and to the prescription
for fixing the effective continuum threshold. The corresponding
errors in the resulting decay constants~are called the {\it
OPE-related error\/} and the {\it systematic error}, respectively.
Let us discuss these in turn.

\vspace{1.5ex} \noindent{\it OPE-related error} \vspace{1ex}

The value of the OPE-related error is obtained as follows: We
perform a bootstrap analysis \cite{bootstrap} by allowing the OPE
parameters to vary over the ranges indicated in (\ref{Table:1}),
using 1000 bootstrap events. Gaussian distributions for all OPE
parameters but $\mu$ are employed. For $\mu$ we assume a uniform
distribution in the corresponding range, which~we~choose to be $1
\leq \mu\;(\mathrm{GeV}) \leq 3$ for charmed mesons and $2 \leq
\mu\;(\mathrm{GeV}) \leq 8$ for beauty mesons. The resulting
distribution of the decay constant turns out to be close to
Gaussian shape. Therefore, the quoted OPE-related error is a
Gaussian~error.

\vspace{1.5ex} \noindent{\it Systematic error} \vspace{1ex}

The systematic error of some hadron parameter determined by the
method of sum rules (i.e., the error related to the intrinsic
limited accuracy of this method) represents the perhaps most
subtle point in the applications of this method. So far no way to
arrive at a {\it rigorous}---in the mathematical
sense---systematic error has been proposed.~Therefore,~in this
respect we have to rely on our experience obtained from the
examples where the exact hadron parameters may be calculated
independently from the method of dispersive sum rules and then
compared with the results of the sum-rule approach. Recent
experience from potential models shows that the band of values
obtained from linear, quadratic,~and cubic Ans\"atze for the
effective threshold encompasses the true value of the decay
constant \cite{lms_new}. Moreover, we could~show that the
extraction procedures in quantum mechanics and in QCD are even
quantitatively rather similar~\cite{lms_qcdvsqm}. Therefore, we
believe that the half-width of this band may be regarded as {\it
realistic\/} estimate for the systematic uncertainty of the decay
constant. Presently, we do not see other possibilities to obtain a
more reliable estimate for the systematic~error.

\subsection{Decay constant of the $D$ meson}
\begin{figure}[!h]
\begin{tabular}{ccc}
\includegraphics[width=5.75cm]{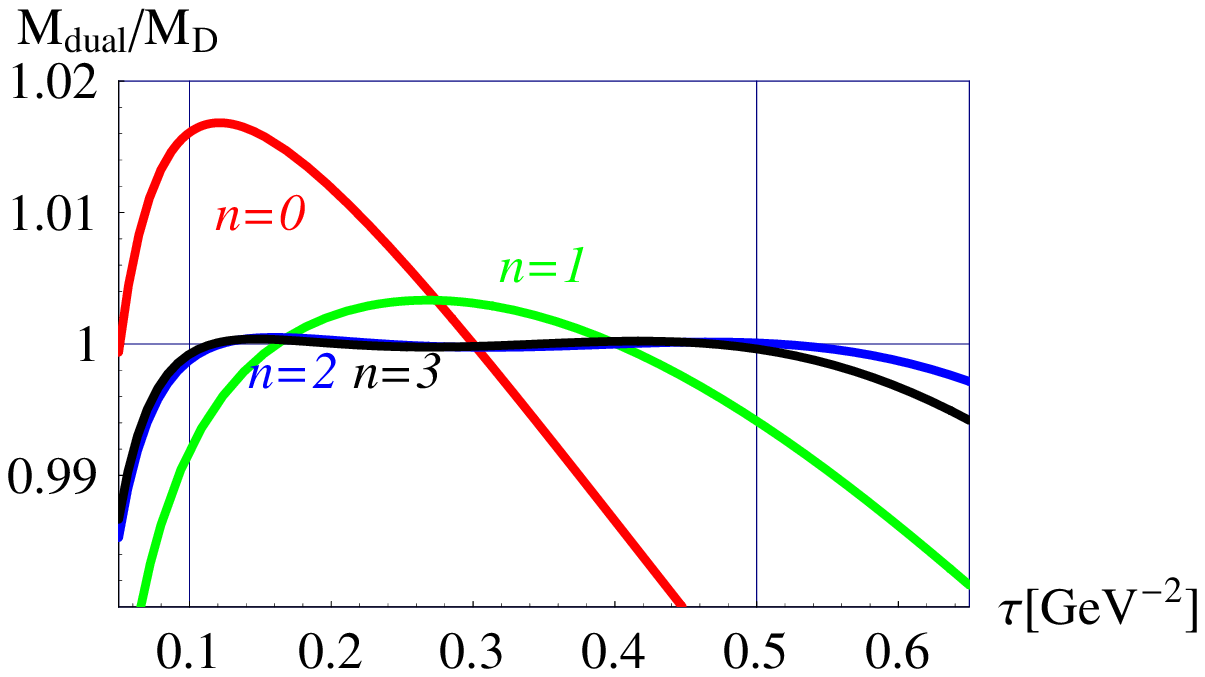}&
\includegraphics[width=5.75cm]{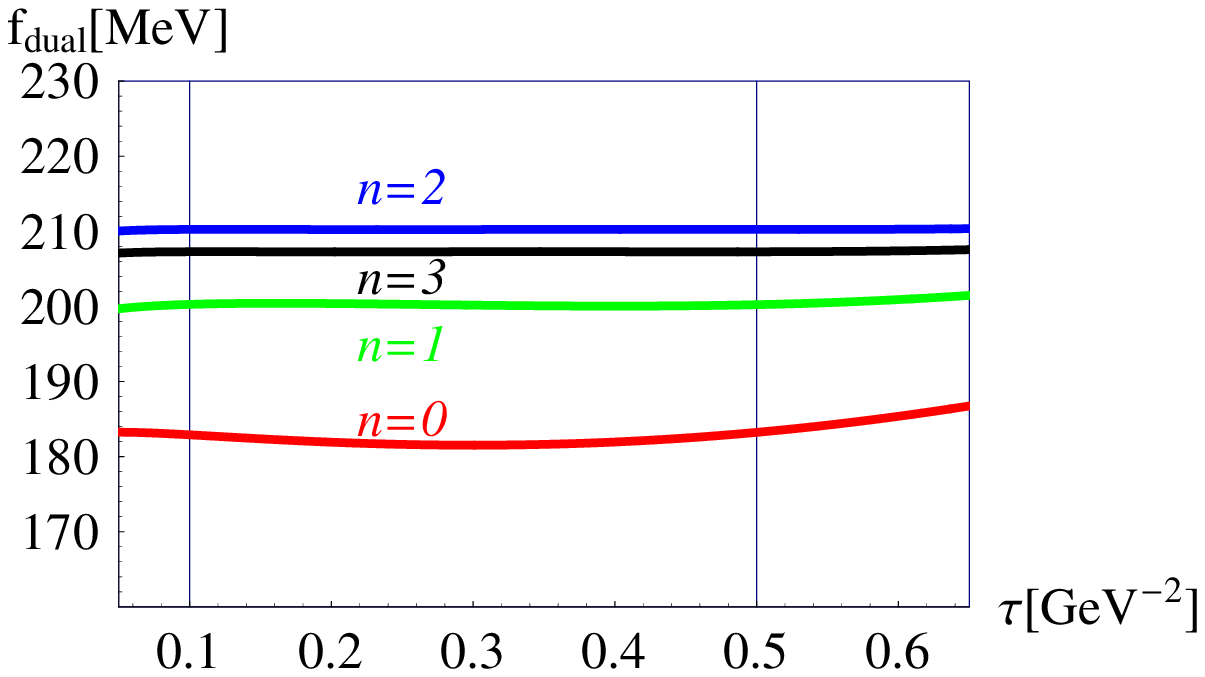}&
\includegraphics[width=5.75cm]{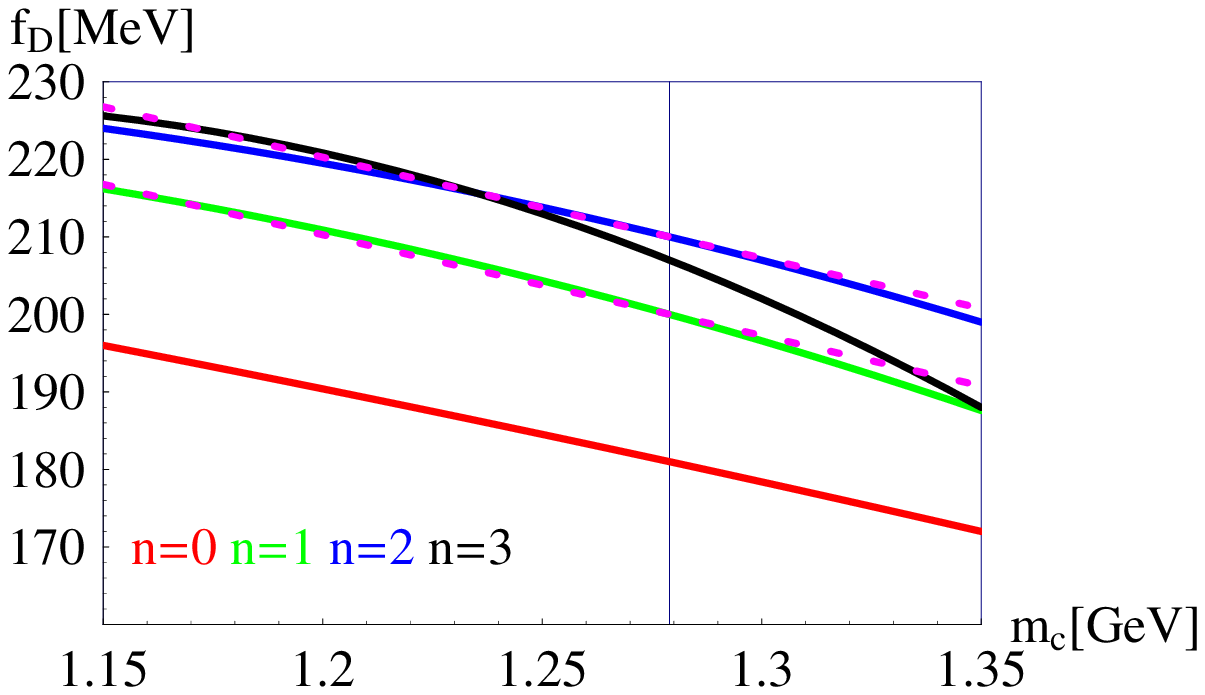}\\
(a) & (b) & (c)
\end{tabular}
\caption{\label{Plot:fD} Dual mass (a) and dual decay constant (b)
of the $D$ meson obtained using different Ans\"atze for the
effective continuum threshold $s_{\rm eff}(\tau)$ (\ref{zeff}) and
fixing all thresholds according to (\ref{chisq}). Results for
$m_c\equiv {\overline m}_c({\overline m_c})=1.279$ GeV, $\mu=m_c$,
and central values of the other relevant parameters are presented.
(c) Dual decay constant of the $D$ meson vs.\ $m_c$ for $\mu=m_c$
and central values of the other OPE parameters. The integer
$n=0,1,2,3$ is the degree of the polynomial in our Ansatz
(\ref{zeff}) for $s_{\rm eff}(\tau)$.}
\end{figure}

The $\tau$-window for the charmed mesons, $\tau=(0.1 - 0.5)\;
\mbox{GeV}^{-2}$, is chosen according to the criteria
formulated~above. Figure~\ref{Plot:fD} shows the application of
our procedure of fixing the effective continuum threshold and
extracting the resulting $f_D$. We would like to point out that,
in the window, the $\tau$-dependent effective thresholds reproduce
the meson mass much better than the constant one
(Fig.~\ref{Plot:fD}a). This signals that the dual correlators
corresponding to the $\tau$-dependent thresholds are less
contaminated by excited states.

The dependence of the extracted value of the $D$-meson decay
constant $f_D$ on the $c$-quark mass $m_c\equiv {\overline
m}_c({\overline m_c})$~and~the condensate $\langle \bar
qq\rangle\equiv \langle \bar qq(2\;{\rm GeV})\rangle$ may be
parameterized as
\begin{equation}
f_{D}^{\rm dual}(m_c,\mu=m_c,\langle \bar qq\rangle) =\left[206.2
-13\left(\frac{m_c-\mbox{1.279\;GeV}}{\mbox{0.1\;GeV}}\right) + 4
\left(\frac{|\langle \bar
qq\rangle|^{1/3}-\mbox{0.267\;GeV}}{\mbox{0.01\;GeV}}\right) \pm
5.1_{\rm (syst)} \right] \mbox{MeV}.
\end{equation}
This formula describes the band of values indicated by the two
dotted lines in Fig.~\ref{Plot:fD}c, which delimit the
results~found from the linear, quadratic, and cubic Ans\"atze for
the effective continuum threshold.
\begin{figure}[!ht]
\begin{tabular}{cc}
\includegraphics[width=5.75cm]{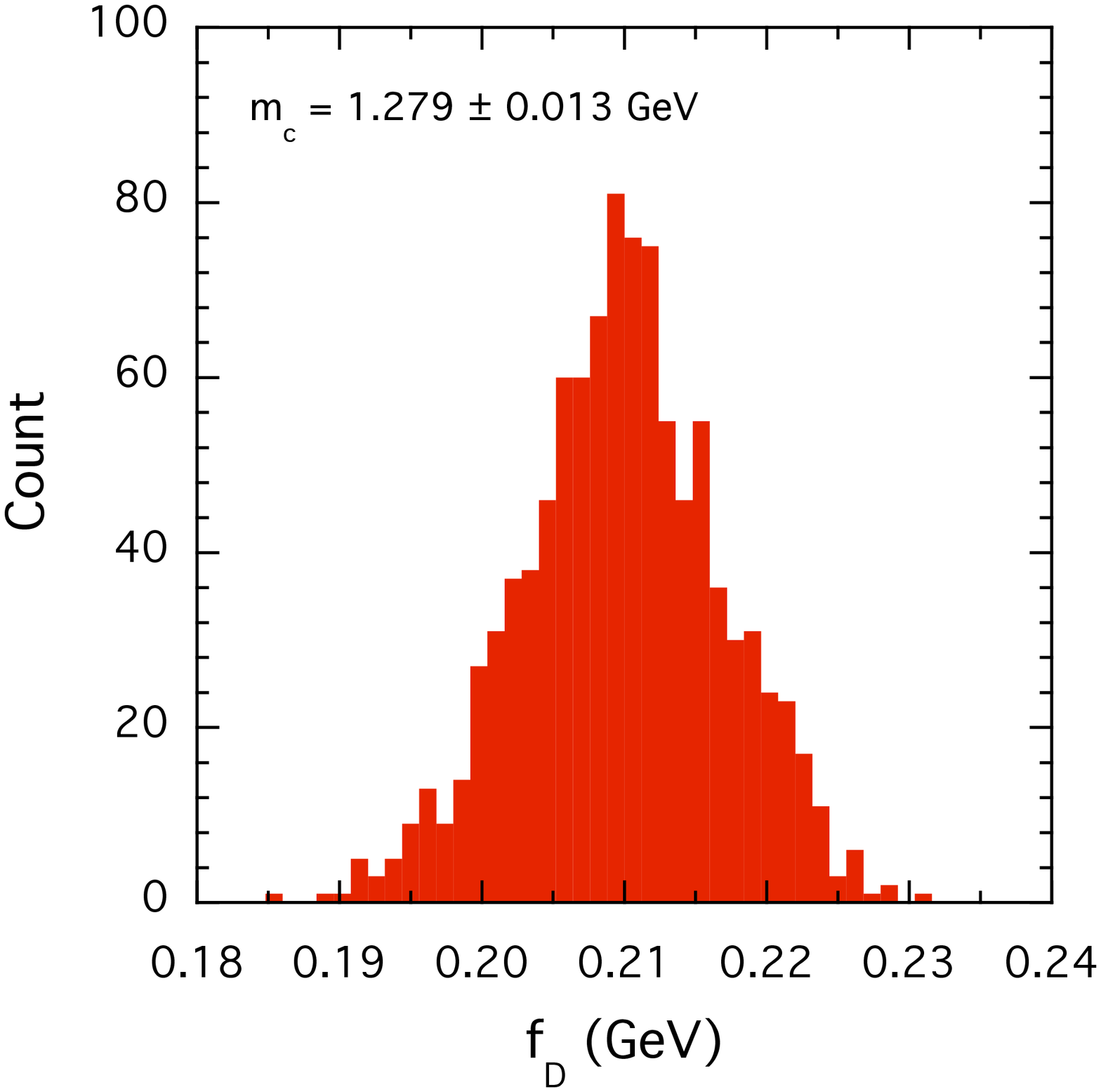}&
\includegraphics[width=5.75cm]{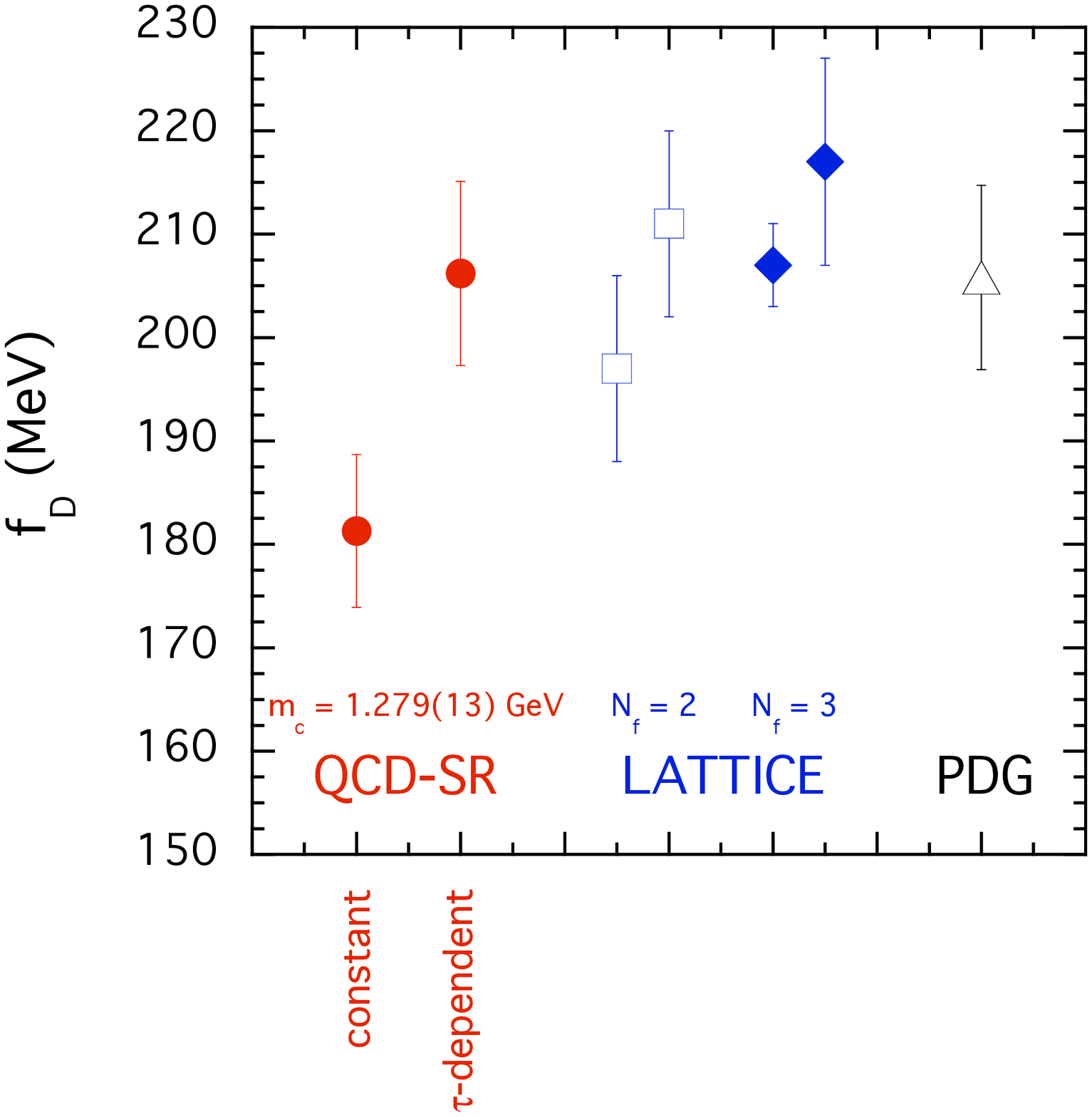}\\
(a) & (b)
\end{tabular}
\caption{\label{Plot:fD_bootstrap} (a) Distribution of $f_D$
obtained by the bootstrap analysis of the OPE uncertainties.
Gaussian distributions for all OPE parameters but $\mu$ with
corresponding errors as given in (\ref{Table:1}) are employed. For
$\mu$ we assume a uniform distribution in the~range $1\;{\rm GeV}
< \mu < 3\;{\rm GeV}$. (b) Summary of findings for $f_D$. Lattice
results are from \cite{ETMC1,ETMC2} for two dynamical
light~flavors ($N_f = 2$) and from \cite{HPQCD1,FNAL+MILC1} for
three dynamical flavors ($N_f=3$). The triangle represents the
experimental value from PDG \cite{pdg}. For the $\tau$-dependent
QCD-SR result the error shown is the sum of the OPE and systematic
uncertainties in (\ref{fD}), added in quadrature.}
\end{figure}
Figure~\ref{Plot:fD_bootstrap}a depicts the result of the
bootstrap analysis of the OPE uncertainties. The distribution has
a Gaussian shape, and therefore the corresponding OPE uncertainty
is the Gaussian error. Adding the half-width of the band deduced
from our $\tau$-dependent Ans\"atze~for the effective continuum
threshold of degree $n=1,2,3$ as the (intrinsic) systematic error,
we obtain the following~result:
\begin{equation}
\label{fD} f_{D}= \left(206.2 \pm 7.3_{\rm (OPE)} \pm 5.1_{\rm
(syst)}\right) \mbox{MeV}.
\end{equation}
The main sources of the OPE uncertainty in the extracted $f_D$ are
its renormalization-scale dependence and the error~of the quark
condensate.

For a $\tau$-independent Ansatz for the effective continuum
threshold a bootstrap analysis entails the substantially~lower
range
%\begin{eqnarray}
%\label{fD_constant}
$f_D^{(n=0)} = \left(181.3\pm 7.4_{\rm (OPE)}\right) \mbox{MeV}$,
%\end{eqnarray}
which differs from our $\tau$-dependent result (\ref{fD}) by
$\simeq 10\%$, i.e.,~by almost~three times the OPE uncertainty.
Moreover, as we have already shown in our previous works
\cite{lms_1}, making use of merely the constant Ansatz for the
effective continuum threshold does not allow one to probe at all
the intrinsic systematic error of the QCD sum rule. From our
result (\ref{fD}) the latter turns out to be of the same order as
the OPE uncertainty.

Allowing the threshold to depend on $\tau$ leads to a clearly
visible effect and brings the results from QCD sum~rules~into
perfect agreement with recent lattice results and the experimental
data (Fig.~\ref{Plot:fD_bootstrap}b). This perfect agreement of
our~result with both experimental data and lattice results
provides a strong argument in favour of the reliability of our
procedure.

\subsection{\boldmath Decay constant of $D_s$ meson}
\begin{figure}[!ht]
\begin{tabular}{ccc}
\includegraphics[width=5.75cm]{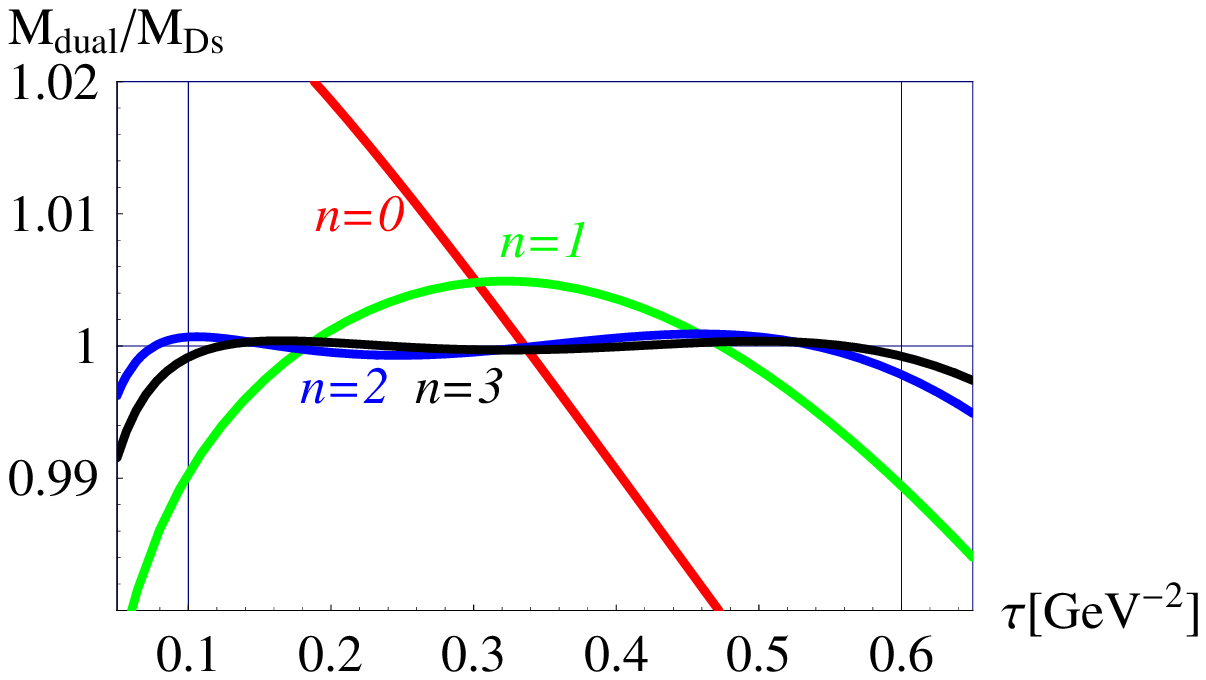}&
\includegraphics[width=5.75cm]{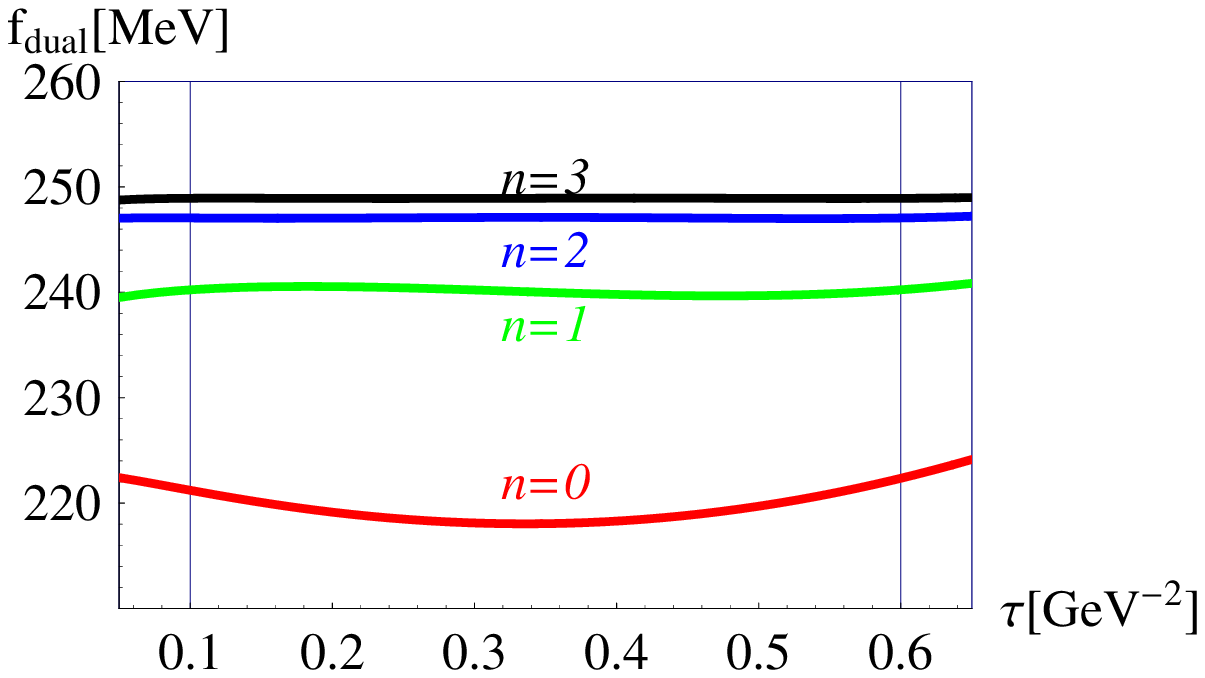}&
\includegraphics[width=5.75cm]{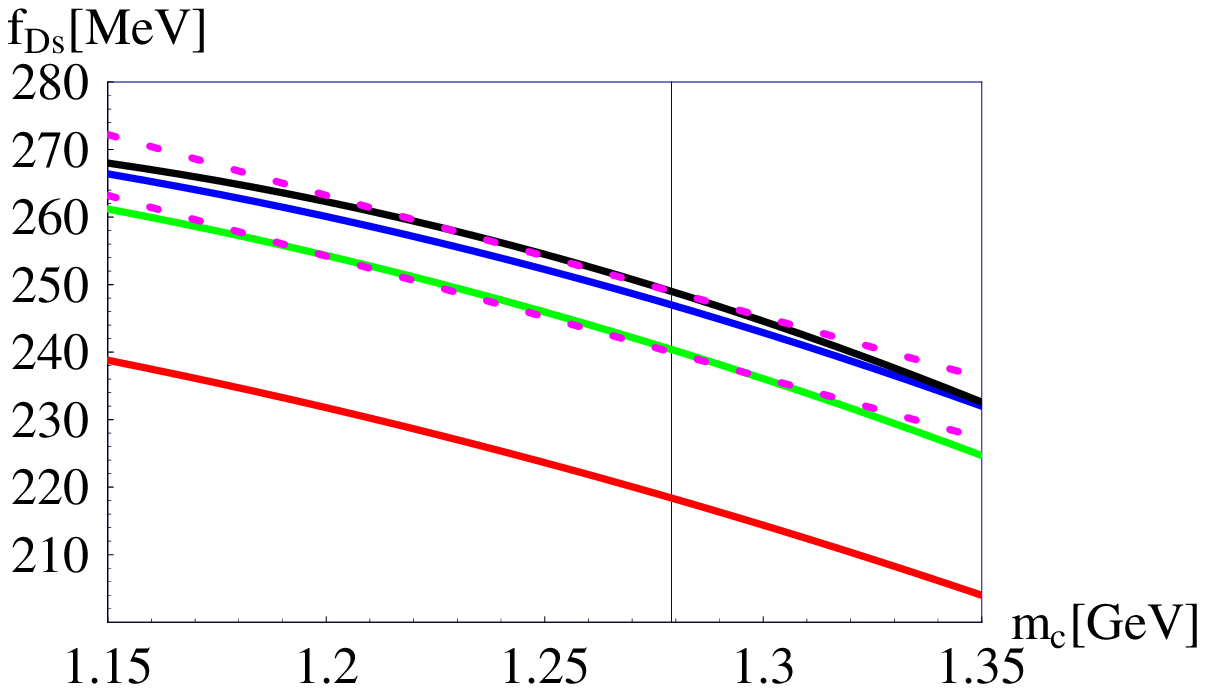}\\
(a) & (b) & (c)
\end{tabular}
\caption{\label{Plot:fDs} Same as Fig.~\ref{Plot:fD} but for the
$D_S$ meson.}
\end{figure}

The corresponding $\tau$-window is $\tau = (0.1-0.6)$\;GeV$^{-2}$.
Figure~\ref{Plot:fDs} provides the details of our extraction
procedure.~Our results for the $D_s$-meson decay constant
$f_{D_s}$ may be represented as
\begin{eqnarray}
f_{D_s}^{\rm dual}(m_c,\mu=m_c,\langle \bar ss\rangle)
=\left[245.3
-18\left(\frac{m_c-\mbox{1.279\;GeV}}{\mbox{0.1\;GeV}}\right) +
3.5 \left(\frac{|\langle \bar
ss\rangle|^{1/3}-\mbox{0.248\;GeV}}{\mbox{0.01\;GeV}}\right) \pm
4.5_{\rm (syst)} \right] \mbox{MeV}.
\end{eqnarray}
This formula describes the band of values indicated by the two
dotted lines in Fig.~\ref{Plot:fDs}c as function of $m_c\equiv
{\overline m}_c({\overline m_c)}$~and gives also the dependence on
the quark condensate $\langle \bar ss\rangle\equiv \langle \bar
ss(2\;{\rm GeV})\rangle$.
\begin{figure}[!ht]
\begin{tabular}{ccc}
\includegraphics[width=5.75cm]{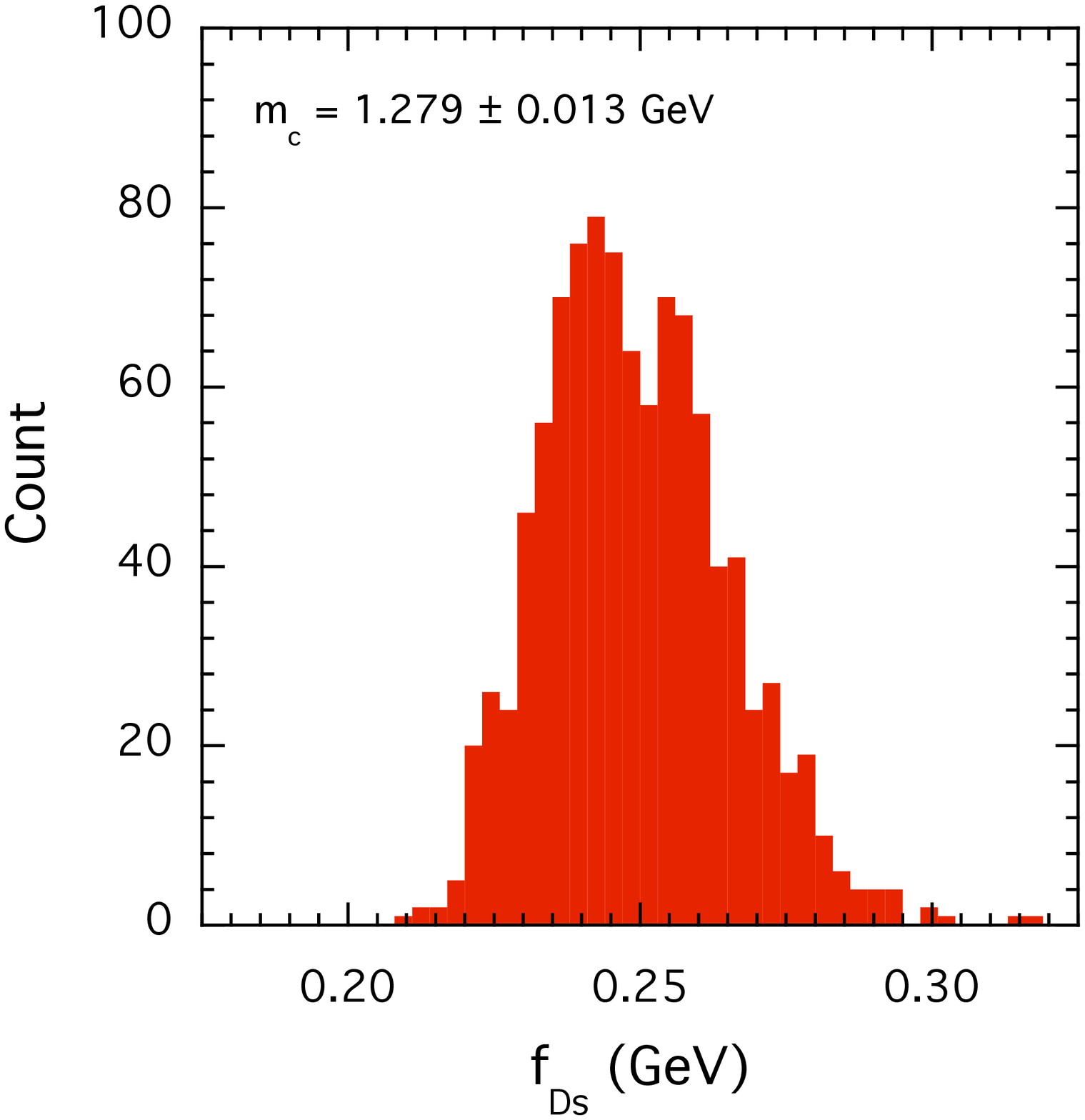}&\qquad
\qquad&\includegraphics[width=5.75cm]{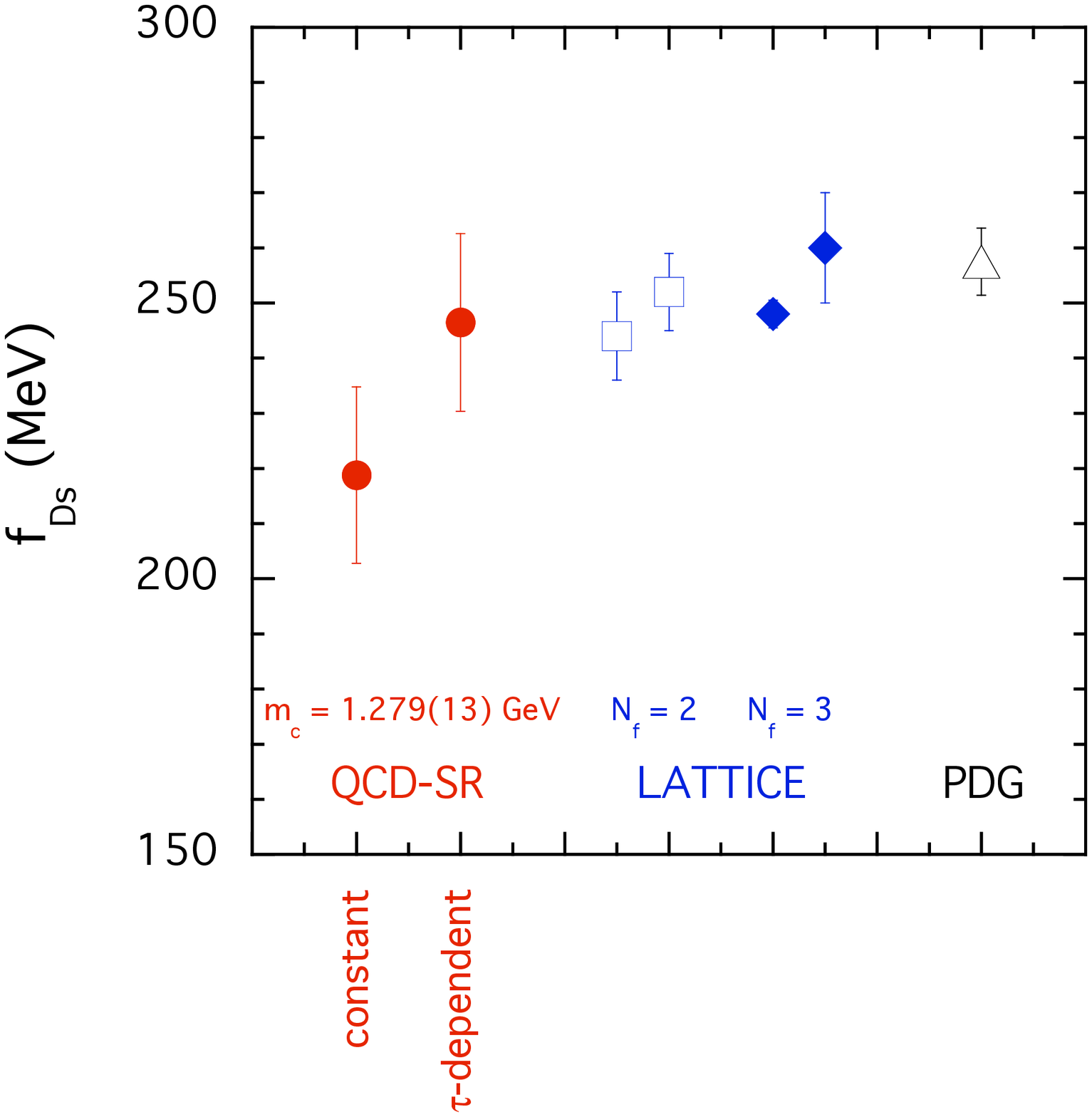}
\\
(a) &\qquad \qquad& (b)
\end{tabular}
\caption{\label{Plot:fDs_bootstrap} (a) Distribution of $f_{D_s}$
obtained by the bootstrap analysis of the OPE uncertainties.
Gaussian distributions for all~OPE parameters but $\mu$ with
corresponding errors as given in (\ref{Table:1}) are employed. For
$\mu$ we assume a uniform distribution in the~range $1\;{\rm GeV}
< \mu < 3\;{\rm GeV}$. (b) Summary of findings for $f_{D_s}$.
Lattice results are from \cite{ETMC1,ETMC2} for two dynamical
light~flavors ($N_f = 2$) and from \cite{HPQCD1,FNAL+MILC1} for
three dynamical flavors ($N_f=3$). The triangle represents the
experimental value from PDG \cite{pdg}. For the $\tau$-dependent
QCD-SR result the error shown is the sum of the OPE and systematic
uncertainties in (\ref{fDs}), added in quadrature.}
\end{figure}
Performing the bootstrap analysis of the~OPE uncertainties, we
obtain the following estimate:
\begin{eqnarray}
\label{fDs} f_{D_s} = \left(245.3 \pm 15.7_{\rm (OPE)} \pm
4.5_{\rm (syst)}\right){\rm MeV}.
\end{eqnarray}
As in the case of $f_D$, a constant-threshold Ansatz yields a
substantially lower value: $f_{D_s}^{(n=0)} = \left(218.8 \pm
16.1_{\rm (OPE)}\right){\rm MeV}$.

\subsection{\boldmath $f_{D_s}/f_D$}
For the ratio of the $D$ and $D_s$ decay constants we report the
sum-rule prediction
\begin{eqnarray}
\label{ratioD}
f_{D_s}/f_D = 1.193\pm 0.025_{(\rm OPE)}\pm 0.007_{(\rm syst)}.
\end{eqnarray}
This value is to be compared with the PDG average $f_{D_s}/f_D =
1.25\pm 0.06$ \cite{pdg} as well as with the recent lattice
results $f_{D_s}/f_D = 1.24 \pm 0.03$ \cite{ETMC1} for $N_f=2$ and
$f_{D_s}/f_D = 1.164 \pm 0.011$ \cite{HPQCD1} and $f_{D_s}/f_D =
1.20 \pm 0.02$ \cite{FNAL+MILC1} for $N_f=3$.~The error in
(\ref{ratioD}) arises mainly from the uncertainties in the quark
condensates $\langle \bar s s\rangle/\langle \bar q
q\rangle=0.8\pm 0.3$.

\section{Summary and conclusions}
We presented a detailed analysis of the decay constants of charmed
heavy mesons with the help of QCD sum rules. Particular emphasis
was laid on the study of the uncertainties in the extracted values
of the decay constants: the~OPE uncertainty related to the not
precisely known QCD parameters and the intrinsic uncertainty of
the sum-rule method related to a limited accuracy of the
extraction procedure.

Our main findings may be summarized as follows.

(i) The perturbative expansion of the two-point function in terms
of the pole mass of the heavy quark exhibits no sign of
convergence. However, reorganizing this expansion in terms of the
corresponding running mass leads to~a clear hierarchy of the
perturbative contributions. Interestingly, the decay constant
extracted from the pole-mass~OPE proves to be sizeably smaller
than the one extracted from the running-mass OPE. In spite of this
numerical difference, the decay constants extracted from these two
correlators exhibit perfect stability in the Borel parameter. This
example shows that stability {\it per se\/} does not guarantee the
reliability of the sum-rule extraction of any bound-state
parameter.

(ii) We have made use of the Borel-parameter-dependent effective
threshold for the extraction of the decay constants. The
$\tau$-dependence of the effective threshold emerges quite
naturally when one attempts to increase the accuracy of the
duality approximation. According to our algorithm, one should
consider different polynomial Ans\"atze for the~effective
threshold and fix the coefficients in these Ans\"atze by
minimizing the deviation of the dual mass from the known~actual
meson mass in the window. Then, the band of values corresponding
to the linear, quadratic, and cubic Ans\"atze reflects the
intrinsic uncertainty of the method of sum rules. The efficiency
of this criterion has been tested before for several examples of
quantum-mechanical models. This strategy has now been applied to
the decay constants of heavy mesons.

(iii) We obtained the following sum-rule estimates for the decay
constants of the charmed $D$ and $D_s$ mesons:
\begin{eqnarray}
\label{fD_final} f_{D}&=& \left(206.2 \pm 7.3_{\rm (OPE)} \pm
5.1_{\rm (syst)}\right) \mbox{MeV}, \\ \label{fDs_final}
f_{D_s}&=& \left(245.3 \pm 15.7_{\rm (OPE)} \pm 4.5_{\rm
(syst)}\right) \mbox{MeV}.
\end{eqnarray}
We point out that we provide both the OPE uncertainties and the
intrinsic (systematic) uncertainty of the method~of sum rules
related to the limited accuracy of the extraction procedure. In
the case of $f_D$ the latter turns out to be~of~the same order as
the OPE uncertainty. Noteworthy, adopting a $\tau$-independent
effective threshold leads to a substantially lower range
$f_D^{(n=0)} = \left(181.3 \pm 7.4_{\rm (OPE)}\right) \mbox{MeV}
$, which differs from our $\tau$-dependent result (\ref{fD_final})
by almost~three~times the OPE uncertainty. The resulting ratio of
the decay constants is
\begin{eqnarray}
\label{ratioD_final}
f_{D_s}/f_D = 1.193\pm 0.025_{(\rm OPE)}\pm 0.007_{(\rm syst)}.
\end{eqnarray}

(iv) Our study of charmed mesons clearly demonstrates that the use
of Borel-parameter-dependent thresholds leads to two essential
improvements:

a. The actual accuracy of the decay constants extracted from sum
rules improves considerably.

b. Our algorithm yields {\it realistic\/} (although not entirely
rigorous) estimates for the systematic errors and allows~one~to
reduce their values to the level of a few percent. Due to the
application of our prescription, the QCD sum-rule results are
brought into perfect agreement both with the experimental results
and with lattice QCD.

\vspace{.5cm} {\bf Acknowledgements.} D.M.\ was supported by the
Austrian Science Fund (FWF), projects no.~P20573 and P22843.

\end{document}